\documentclass[twocolumn,showpacs,secnumarabic,amssymb, aps, prb]{revtex4-1}

\usepackage[version=3]{mhchem} 
\usepackage{bibentry,natbib}
\usepackage{graphicx}
\usepackage{notes2bib}
\begin{document}

\title{Structure and thermodynamic properties of a weakly-coupled antiferromagnetic spin-1/2 chain compound \ce{(C_5H_{12}N)CuBr_3}}%

\author{B. Y. Pan,$^1$ Y. Wang,$^2$ L. J. Zhang,$^{2,*}$ and S. Y. Li$^{1,2,*}$}%
\affiliation{$^1$State Key Laboratory of Surface Physics, Department of
Physics, Fudan University, Shanghai 200433, P. R. China}
\affiliation{$^2$Laboratory of Advanced Materials, Fudan University,
Shanghai 200438, P. R. China}
\date{\today}

\begin{abstract}
  Single crystals of a metal organic complex \ce{(C5H12N)CuBr3} (\ce{C5H12N} = piperidinium, pipH for short) have been synthesized and the structure was determined by single-crystal X-ray diffraction. \ce{(pipH)CuBr3} crystallizes in the monoclinic group $C$2/$c$. Edging-sharing \ce{CuBr5} units link to form zigzag chains along the $c$ axis and the neighboring Cu(II) ions with spin-1/2 are bridged by bi-bromide ions. Magnetic susceptibility data down to 1.8 K can be well fitted by the Bonner-Fisher formula for antiferromagnetic spin-1/2 chain, giving the intrachain magnetic coupling constant $J$ $\sim$ 17 K. At zero field, \ce{(pipH)CuBr3} shows three-dimensional (3D) order below $T_N$ = 1.68 K. Calculated by the mean-field theory, the interchain coupling constant $J'$ = 0.65 K is obtained and the ordered magnetic moment $m_0$ is about 0.20 $\mu_B$. This value of $m_0$ makes \ce{(pipH)CuBr3} a rare compound suitable to study the dimensional crossover problem in magnetism, since both 3D order and one-dimensional (1D) quantum fluctuations are prominent. In addition, specific heat measurements reveal two successive magnetic transitions with lowering temperature when external field $H \geq$ 3 T is applied along the $a'$ axis. The $H$ - $T$ phase diagram of \ce{(pipH)CuBr3} is roughly constructed. The interplay between exchange interactions, dimensionality, Zeeman energy and possible Dzyaloshinkii-Moriya interaction should be the driving force for the multiple phase transitions.
\end{abstract}


\maketitle

\section{Introduction}
The conventional 3D magnets are described by the semi-classical spin wave theory, whereas the 1D magnets are described fully by quantum theory.\cite{Majlis2007} What in particular interest is the Heisenberg antiferromagnetic (HAFM) spin chain which represents one of the few exactly solvable many-body models in quantum physics.\cite{Bethe1931,Huthen1938} The HAFM spin chain is a disordered system with fractional excitations called spinons.\cite{Tennant1995} Its multispinon continuum spectrum is distinct from the sharp spin wave spectrum in conventional 3D magnets.\cite{Tennant1995} In real compounds, the weak interchain interaction $J'$ between HAFM spin chains will lead the system to a 3D ordered state at sufficiently low temperature. Such a 3D ordered state is essentially different from that of conventional 3D magnets, since its spectrum has both sharp spin waves at low energies and multispinon continuum at high energies.\cite{Coldea2001,Coldea2003,Kohno2007,Zheludev2000,Zheludev2002,Lake2005} In this sense, the 3D ordered state of weakly-coupled HAFM spin chains lies in the crossover regime from quantum to semi-classical physics. A novel longitudinal spin wave mode, which does not exist in conventional 3D magnets, is predicted to emerge by mean-field and random phase approximation theories (MF/RPA).\cite{Schulz1996,Essler1997} However, the results of inelastic neutron scattering experiments showed substantial discrepancies with theories, which was attributed to the ignorance of correlation effects in MF/RPA.\cite{Zheludev2002,Lake2000,Lake2005,Zheludev2003} Therefore, the dimensional crossover problem remains to be solved in magnetism.\cite{Zhitomirsky2013,Schulz1996,Zheludev2002,Kohno2007}

The real compounds of weakly-coupled HAFM spin chains can be classified into two categories. The first one is copper(II) based inorganic compounds such as \ce{BaCu2Si2O7},\cite{Tsukada1999} \ce{SrCuO2} and \ce{Sr2CuO3},\cite{Motoyama1996} \ce{Ca2CuO3},\cite{Kiryukhin2001} and \ce{KCuF3},\cite{Hirakawa1970} in which $J$ is in the order of 1000 K. The second one is 1D molecular magnets in which $J$ is usually below 100 K.\cite{Landee2013} The 1D molecular magnets are mostly metal-organic complexes such as copper benzoate,\cite{Date1970} copper pyrimidine,\cite{Ishida1997,Feyerherm2000} copper pyrazine dinitrate,\cite{Hammar1999} and \ce{CuCl2\cdot2((CH3)2SO)},\cite{Chen2007} in which the spins of neighboring Cu(II) ions interact via water molecules, halide ions, or bridging ligands.\cite{Landee2013} The ordered state of weakly-coupled HAFM spin chains has dual nature of 1D and 3D characters. In order to study the dimensional crossover problem, the ordered magnetic moment $m_0$ should have a proper value, so that both the 1D and 3D characters are prominent. However, real compounds that meet this condition are rare. To our knowledge, \ce{BaCu2Si2O7} with $m_0$ = 0.16 $\mu_B$ is one suitable compound, \cite{Kenzelmann2001,Zheludev2000,Zheludev2002,Zheludev2003,Casola2012} while the $m_0$ of other compounds is either too small or too large.\cite{Kenzelmann2001,Chen2007,Hutchings1969,Kojima1997} Due to the lack of satisfactory theory of the dimensional crossover problem, more compounds with $m_0$ close to that of \ce{BaCu2Si2O7} are highly desirable.

Here, we report the synthesis, structure, and thermodynamic properties of \ce{(pipH)CuBr3}.\bibnote{In a previous literature of spin ladder compound \ce{(pipH)2CuBr4} (B. R. Patyal, B. L. Scott, and R. D. Willett,  Phys. Rev. B \textbf{41}, 1657 (1990)), the authors also mentioned the produce of \ce{(pipH)CuBr3}, but no structure information or properties of this compound were reported.} It is shown that \ce{(pipH)CuBr3} is a weakly-coupled HAFM spin-1/2 chain compound with $J$ $\sim$ 17 K and $J'$ = 0.65 K. Below $T_N$ = 1.68 K, the ordered magnetic moment of Cu(II) ions is $m_0$ = 0.20 $\mu_B$, making \ce{(pipH)CuBr3} another ideal compound to investigate the dimensional crossover problem in magnetism.
\section{Experiment}
Single crystals of \ce{(pipH)CuBr3} were synthesized by evaporation method in a solution of ethanol. 8.93 g Copper(II) bromide was dissolved in 225 ml of ethanol. 2.93 ml aqueous HBr (48\% weight) was added to 1.98 ml piperidine resulting in piperidinium bromide (pipHBr). Then the solution of pipHBr was added slowly to the \ce{CuBr2} solution. After one week of slow evaporation, black single crystals formed at the bottom of the beaker. The solution was decanted and single crystals were harvested. The typical size of the crystals is 3 mm $\times$ 2 mm $\times$ 1.5 mm.

X-ray diffraction of \ce{(pipH)CuBr3} single crystal for the structure determination was carried out on a Bruker SMART Apex (II) diffractometer (Mo $K_\alpha$ radiation, $\lambda$ = 0.71073 \AA). The crystal structure was solved by the direct method and refined via full-matrix least-square techniques using the SHELXL-97 program package.\cite{Sheldrick2008} The results of cell parameters are listed in Table 1. The crystallographic data has been deposited at the Cambridge Crystallographic Data Center (CCDC 970073).

Magnetic susceptibility of \ce{(pipH)CuBr3} single crystal as well as powdered sample were measured by a MPMS SQUID magnetometer (Quantum Design).  For the single crystal, magnetic fields were applied along the $b$ axis, the $c$ axis, and perpendicular to the $bc$ plane, respectively. In the following, we define the direction perpendicular to the $bc$ plane as the $a'$ axis. Specific heat of \ce{(pipH)CuBr3} single crystals was measured by the two-tau method in a small dilution refrigerator integrated to a Physical Property Measurement System (PPMS, Quantum Design) with magnetic fields applied along the $a'$ and $c$ axes.

\section{Results and Discussion}
\begin{figure}
  \includegraphics[clip,width=8.5cm]{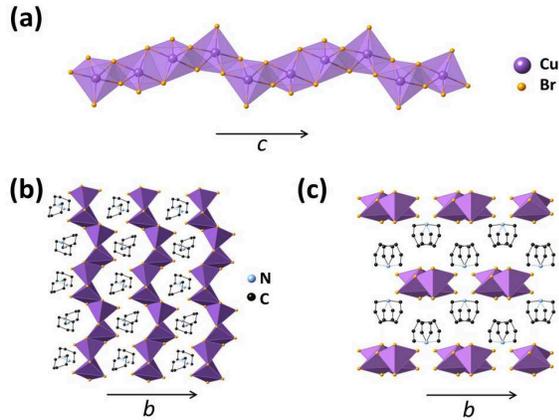}
  \caption{Crystal structure of \ce{(pipH)CuBr3}. (a) Illustration of a single zigzag chain of edge-sharing \ce{CuBr5} units running along the $c$ axis. (b) Packing diagram of the chains on the $bc$ plane. Hydrogen atoms are omitted for clarity.  (c) Crystal structure as viewed along the $c$ axis. Chains are well isolated by the piperidinium cations and show ABA stacking order along the $a'$ direction.}
\end{figure}
\begin{table}
  \caption{Crystal data and structure refinement for \ce{(pipH)CuBr3}.}
  \label{tbl:example}
  \begin{tabular}{ll}
    \hline
    Empirical formula & \ce{C5H12NCuBr3}\\
    Formula weight & 389.43 \\
    Temperature & 296(2) K \\
    Wavelength & 0.71073 \AA \\
    Crystal system & Monoclinic \\
    space group & $C$2/$c$\\
    Unit cell dimensions &  $a$ = 18.906(5) \AA,   $\alpha$ = 90$^\circ$\\
                         &  $b$ = 8.834(3) \AA,    $\beta$ = 103.995(3)$^\circ$\\
                         &  $c$ = 12.665(4) \AA,   $\gamma$ = 90$^\circ$\\
    Volume & 2052.6(10) \AA$^3$\\
    $Z$ & 8\\
    Calculated density & 2.520 Mg/m$^3$\\
    Absorption coefficient & 13.745 mm$^{-1}$\\
    $F$(000) & 1464\\
    Crystal size & 0.3 $\times$ 0.2 $\times$ 0.15 mm\\
    Theta range for data collection & 2.22 to 25.01$^\circ$\\
    Limiting indices & -21 $\leq$ h $\leq$ 22,\\
                     & -9 $\leq$ k $\leq$ 10,\\
                     & -14 $\leq$ l $\leq$ 15\\
    Reflections collected/unique &  5645/1809 [$R_{\mathrm{int}}$ = 0.0456]\\
    Completeness to $\theta$ = 25.01$^\circ$  &  99.7\%\\
    Absorption correction & Semi-empirical from equivalents\\
    Max. and min. transmission & 0.7457 and 0.2959\\
    Refinement method & Full-matrix least-squares on $F^2$\\
    Data/restraints/parameters & 1809/0/92\\
    Goodness-of-fit on $F^2$ & 1.175\\
    Final $R$ indices [$I > 2\sigma(I)$] & $R$1 = 0.0296, $wR$2 = 0.0711\\
    $R$ indices (all data) & $R$1 = 0.0365, $wR$2 = 0.0811\\
    Extinction coefficient & 0.0081(3)\\
    Largest diff. peak and hole & 0.916 and -0.580 e$\cdot$\AA$^{-3}$\\
    \hline
  \end{tabular}
\end{table}

The compound \ce{(pipH)CuBr3} crystallizes in the monoclinic space group $C$2/$c$ with lattice parameters $a$ = 18.906(5) \AA, $b$ = 8.834(3) \AA, $c$ = 12.665(4) \AA{}, and $\beta$ = 103.995(3)$^\circ$. Figure 1(a) shows the edging-sharing \ce{CuBr5} units linking to form zigzag chains along the $c$ axis with four inequivalent Cu(II) sites in each chain unit. In the square-pyramidal \ce{CuBr5} unit, Cu(II) ion is five-coordinated. The axial bond is 2.8022 \AA{}  and the basal bonds range from 2.4119 to 2.4628 \AA{}. Neighboring Cu(II) ions with spin-1/2 along the chain are bridged by bi-bromide ions. This kind of bi-halide bridged magnetic chains with five-coordination Cu(II) ion can also be found in cyclopentylammonium trichlorocuprate(II), cyclohexylammonium trichlorocuprate(II), $n$-methyl-2-aminopyridiniumtrichlorocuprate(II) ($n$ = 4, 6),\cite{Geiser1986} and \ce{[NO2BzMePy][CuCl3]}.\cite{Han2012,Landee2013} Among these five coordinative halides, four serve as bridging ligands and the fifth as terminal. Along the $b$ direction, chains are linked by hydrogen bonds, as shown in Fig. 1(b). In Fig. 1(c), we show the crystal structure as viewed along the $c$ axis. The chains are well isolated by the piperidinium cations and show ABA stacking order along the $a'$ direction. The crystal data is listed in Table 1.
\begin{figure}
  \includegraphics[clip,width=6.45cm]{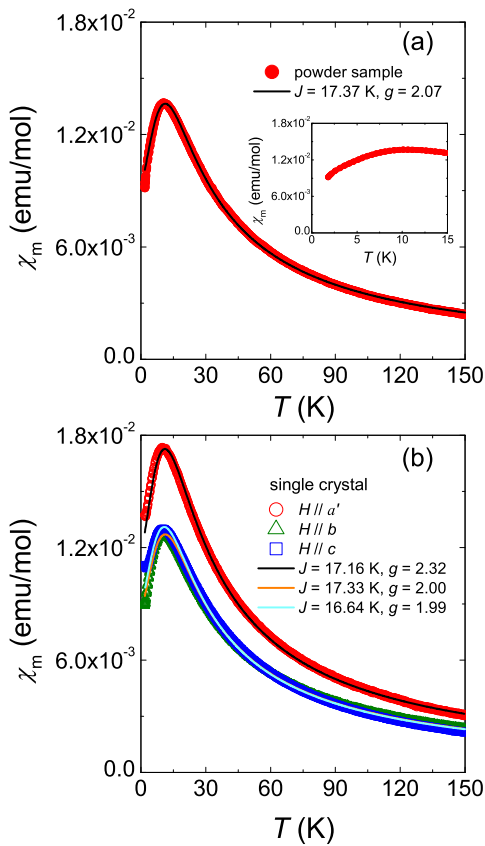}
  \caption{Magnetic susceptibility of \ce{(pipH)CuBr3} as a function of temperature. (a) Data of powder sample. The line is the fitting by Bonner-Fisher formula for data from 4 to 150 K. Inset: susceptibility curve below 15 K  which has a broad maximum around 10 K and a nonzero extrapolation value to zero temperature. (b) Data of single crystal with fields applied along the $a'$ (red circle), $b$ (olive triangle), and $c$ (blue square) axes. Each curve is fitted by Bonner-Fisher formula for data from 4 to 150 K. The fitting parameters are listed in the figure and text.}
\end{figure}

The 1D magnetism of \ce{(pipH)CuBr3} has been confirmed by magnetic susceptibility and specific heat measurements, as will be presented below. Magnetic susceptibility of \ce{(pipH)CuBr3} powder was measured from 1.8 to 150 K in an applied field of 1000 Oe, and the results are shown in Fig. 2(a). The powder was obtained by grinding single crystals. The data show the characteristic feature of HAFM spin-1/2 chain with a broad maximum at about 10 K and a nonzero extrapolation value down to zero temperature,\cite{Dagotto1996} as shown in the inset of Fig. 2(a). The susceptibility data from 4 to 150 K can be well fitted by the Bonner-Fisher formula for HAFM spin-1/2 chain.\cite{Bonner1964} The line in Fig. 2(a) is the best fit with intrachain exchange interaction $J$ = 17.37 $\pm$ 0.01 K and Lander factor $g$ = 2.07. Magnetic susceptibility of \ce{(pipH)CuBr3} single crystal with field of 1000 Oe applied along the $a'$, $b$, and $c$ axes are shown in Fig. 2(b). Again, all three curves can be fitted by the Bonner-Fisher formula from 4 to 150 K, and we get the parameters $J_{a'}$ = 17.16 $\pm$ 0.02 K, $g_{a'}$ = 2.32, $J_b$ = 17.33 $\pm$ 0.01 K, $g_b$ = 2.00, and $J_c$ = 16.64 $\pm$ 0.02 K, $g_c$ = 2.00, respectively. Therefore, the magnetic susceptibility behaviors strongly suggest that \ce{(pipH)CuBr3} is a highly isotropic 1D HAFM spin-1/2 system. 

\begin{figure}
  \includegraphics[clip,width=6.23cm]{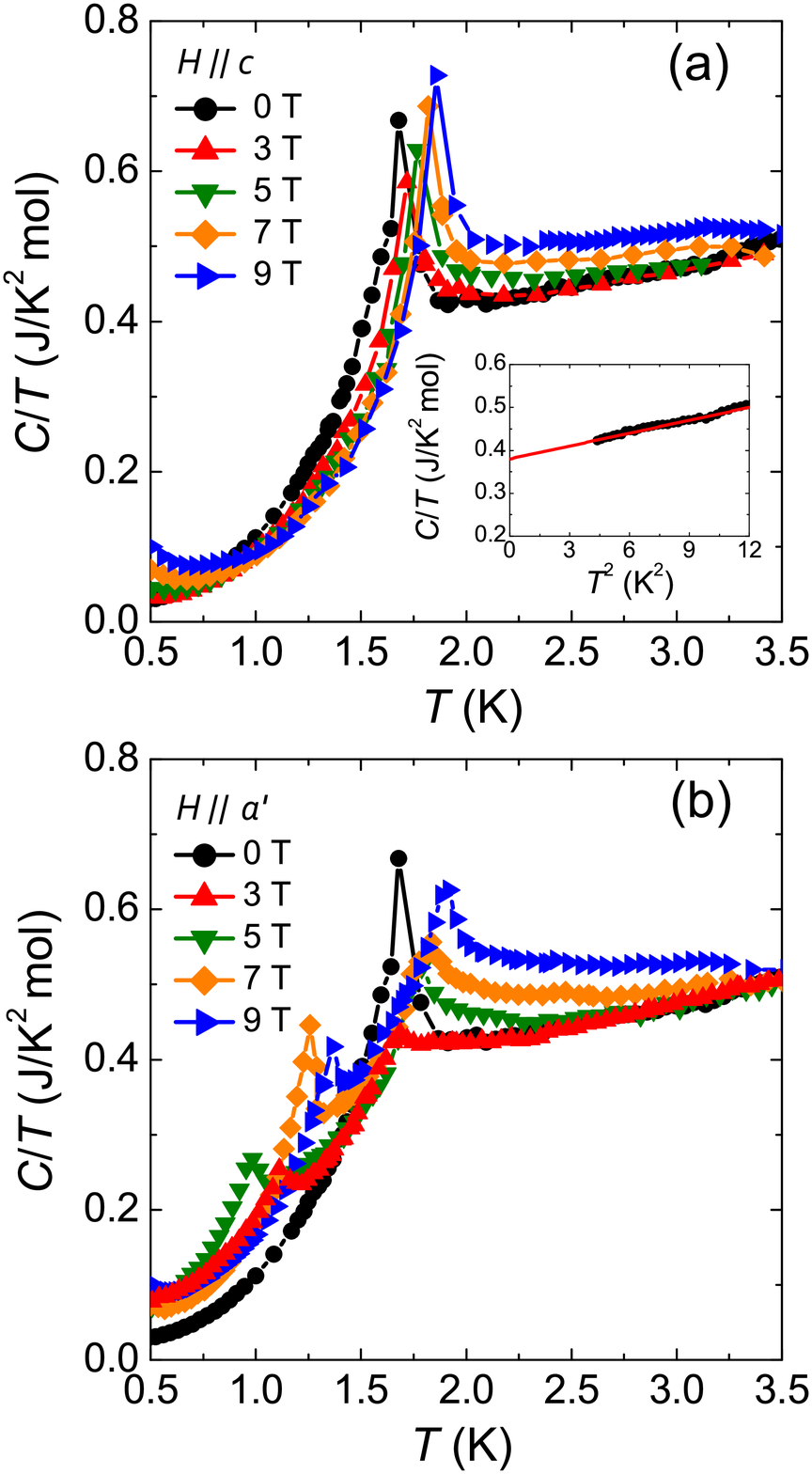}
  \caption{Low temperature specific heat of \ce{(pipH)CuBr3} single crystal. (a) Specific heat with fields applied along the $c$ axis. Inset: zero field specific heat from 2 to 3.5 K in which $C/T$ is plotted as a function of $T^2$. The solid line is the fitting curve $C/T$ = 0.38 + 0.01$T^2$. (b) Specific heat with fields applied along the $a'$ axis.}
\end{figure}

\begin{figure}
  \includegraphics[clip,width=6cm]{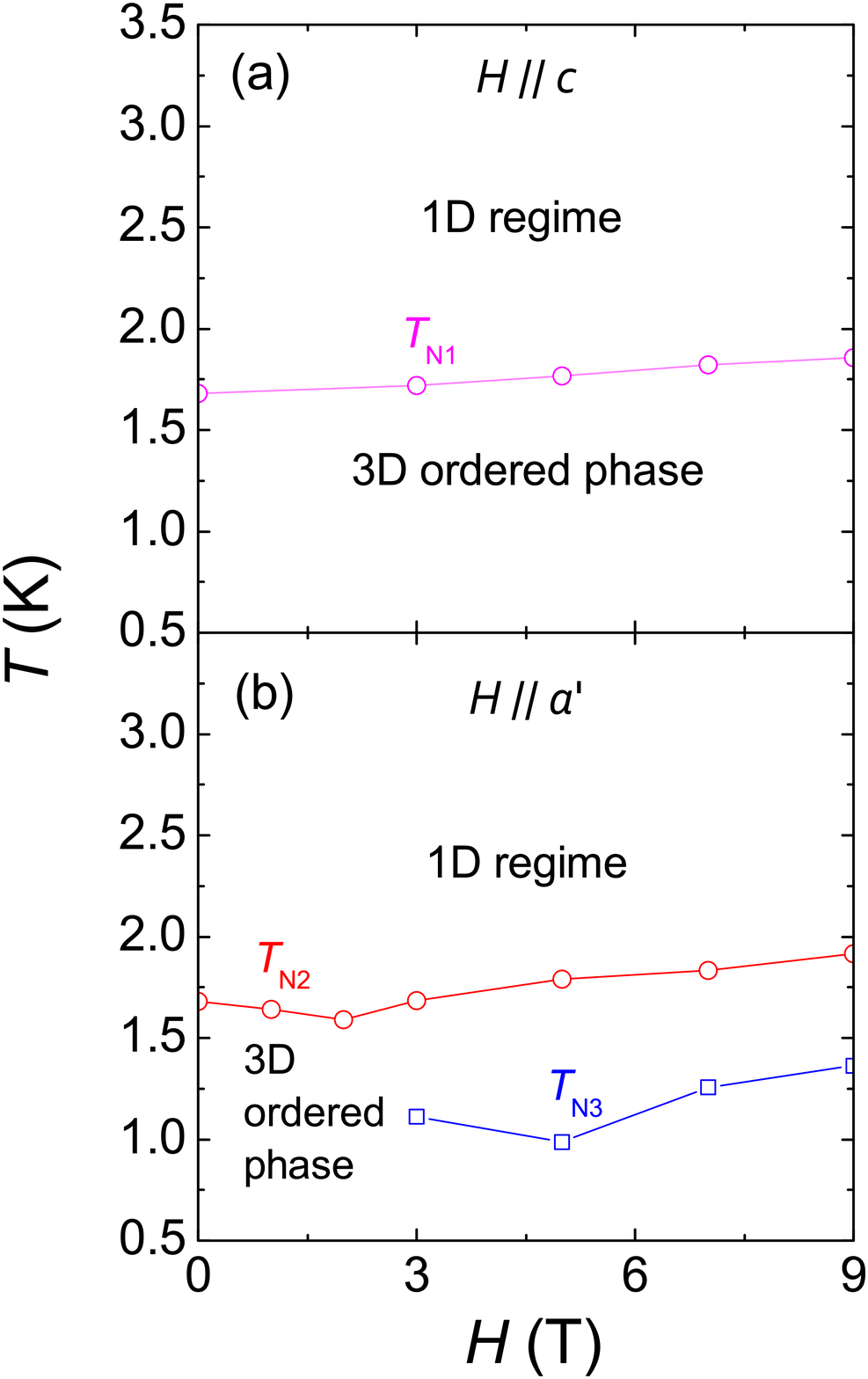}
  \caption{$H$ - $T$ phase diagram of \ce{(pipH)CuBr3} derived from the specific heat measurement. The data points correspond to the peaks of the specific heat curves in Fig. 3.}
\end{figure}
The specific heat of \ce{(pipH)CuBr3} single crystals from 0.5 to 3.5 K are shown in Fig. 3. The zero field data from 2 to 3.5 K can be fitted by the equation \[ C/T = A + BT^2 ,\] as seen in the inset of Fig. 3(a) in which $C/T$ is plotted as a function of $T^2$. The first term represents the 1D spinon contribution and the second term is the phonon contribution. The fitting parameters are $A$ = 0.38 J/K$^2$ mol and $B$ = 0.01 J/K$^4$ mol. Theoretically, the specific heat contributed by spinons of HAFM spin-1/2 chain can be calculated from the equation\cite{Bonner1964}\[ C_S = \frac{2Nk_B^2}{3J}T, \] where $N$ is the Avogadro's constant and $k_B$ is the Boltzmann constant. Taking $J$ = 17.37 K, the above equation gives the linear specific heat coefficient 0.32 J/K$^2$ mol, which is close to our experimental value 0.38 J/K$^2$ mol. In this sense, the specific heat result further confirms the 1D magnetism in \ce{(pipH)CuBr3}. The linear specific heat contribution of spinons has also been observed in other HAFM spin-1/2 chain systems such as copper benzoate,\cite{Dender1997} copper pyrimidine,\cite{Feyerherm2000} and copper pyrazine dinitrate.\cite{Hammar1999}

At zero field, $C(T)$ shows a sharp peak at $T$ = 1.68 K, suggesting the onset of 3D ordered state. The transition temperature $T_N$ varies with the strength and directions of the applied fields. For fields applied along the $c$ axis, the ordering temperature $T_N$ increases slightly with increasing field, as shown in Fig. 3(a). More interestingly, when the field is applied along the $a'$ axis, two peaks are clearly observed for $H \geq$ 3 T, indicating two successive magnetic phase transitions, as shown in Fig. 3(b). For $H <$ 3 T, the low-temperature peak cannot be resolved. We plot a rough $H$ - $T$ phase diagram of \ce{(pipH)CuBr3} in Fig. 4. The data points correspond to the peaks of the specific heat curves in Fig. 3, marked as $T_{N1}$, $T_{N2}$, and $T_{N3}$. The field can easily modulate the magnetic phase, which indicates that the Zeeman term $H_Z = g\mu_BH\displaystyle{\sum_{i}}S_i^Z$ is important in the spin Hamiltonian \[ H_S = {\sum_{i}\{J{\bf{S}}_i\cdot{\bf{S}}_{i+1}} + J'\sum_{\delta_\perp}{\bf{S}}_i\cdot{\bf{S}}_{i,\delta_\perp} + g\mu_BHS_i^Z\}. \] This is reasonable since the Zeeman energy is comparable with the exchange energy in our field range. The phenomenon of field-induced successive magnetic phase transitions has previously been observed in quantum antiferromagnets such as \ce{Cs2CuCl4},\cite{Tokiwa2006} which was attributed to the competition between exchange interactions, Zeeman energy, quantum fluctuations, and Dzyaloshinsky-Moriya (D-M) interaction.\cite{Tokiwa2006} In our \ce{(pipH)CuBr3}, each chain unit has four inequivalent Cu(II) ions, and this alternating structure would induce staggered D-M interaction. Since we cannot determine the magnetic structures in zero and magnetic fields, the exact magnetic phase diagram of \ce{(pipH)CuBr3} needs to be revealed by other experiments such as neutron scatting measurement.

The 3D ordered state of weakly-coupled HAFM spin-1/2 chains is rather exotic in which low-energy spin waves coexist with high-energy multispinon continuum. From the ordering temperature $T_N$, the interchain exchange interaction $J'$ and the ordered moment $m_0$ can be calculated by the mean field theory:\cite{Schulz1996}
 \[\mid J'\mid = \frac{T_N}{1.28\sqrt{\mathrm{ln}(5.8J/T_N)}},\]  \[m_0 = 1.017\sqrt{\frac{J'}{J}}.\]
Taking $T_N$ = 1.68 K at zero field and $J$ = 17.37 K, the above equations give $J'$ = 0.65 K and $m_0$ = 0.20 $\mu_B$. This value of $m_0$ in the ordered state indicates 80\% reduction from the saturated moment 1 $\mu_B$ of Cu(II) ions due to quantum fluctuations.

Experimentally, a weakly-coupled HAFM spin-1/2 chain system suitable to study the dimensional crossover problem should has a proper $m_0$ value. If $m_0$ is too small, as in \ce{Sr2CuO3} ($\sim$ 0.06 $\mu_B$) and \ce{Ca2CuO3} ($\sim$ 0.09 $\mu_B$),\cite{Kojima1997} the 1D character still dominates below the 3D ordered temperature and spin waves are hard to identify in the spectrum.\cite{Zheludev2000,Kojima1997} On the other hand, if $m_0$ is large, as in \ce{KCuF3} ($\sim$ 0.54 $\mu_B$),\cite{Hutchings1969} the spin waves are very intense and the multispinon continuum is hard to isolate.\cite{Zheludev2000,Tennant1995-2} So far, \ce{BaCu2Si2O7} with $m_0$ = 0.16 $\mu_B$ is an ideal compound,\cite{Tsukada1999} in which both spin waves and multispinon continuum have been clearly observed by inelastic neutron scattering measurements.\cite{Zheludev2000} However, the experimentally obtained excitation spectrum of \ce{BaCu2Si2O7} have serious discrepancies with theories on issues such as the dynamic structure factors of the novel
longitudinal spin wave mode.\cite{Zheludev2002,Zheludev2003} The $m_0$ of \ce{(pipH)CuBr3} is close to that of \ce{BaCu2Si2O7}. Further studying the excitation spectrum of \ce{(pipH)CuBr3} may help to resolve the dimensional crossover problem in magnetism.
\section{Summary}
In summary, we report the synthesis, structure, and thermodynamic properties of a new quantum 1D magnetic compound \ce{(pipH)CuBr3}. The crystal shows 1D structure of Cu(II) ions connected by bi-bromide bridge along the $c$ axis. From the magnetic susceptibility and specific heat measurements, we demonstrate that \ce{(pipH)CuBr3} can be well described by weakly-coupled HAFM spin-1/2 chains with intrachain interaction $J$ $\sim$ 17 K and interchain interaction $J'$ = 0.65 K. The proper value of $m_0$ = 0.20 $\mu_B$ in the 3D ordered state at zero field makes it another ideal compound to study the dimensional crossover problem in magnetism, except for \ce{BaCu2Si2O7}. The observation of two successive magnetic transitions in specific heat measurements indicates a complex magnetic phase diagram of \ce{(pipH)CuBr3}, which is roughly constructed.

\begin{center}
{\bf ACKNOWLEDGEMENTS}
\end{center}
We thank Y. Chen for helpful discussions. This work is supported by the Natural Science Foundation of China, the Ministry of Science and Technology of China (National Basic Research Program No: 2012CB821402), Program for Professor of Special Appointment (Eastern Scholar) at Shanghai Institutions of Higher Learning.\\

$^*$zhanglijuan@fudan.edu.cn; shiyan$\_$li@fudan.edu.cn
\bibliography{pipHCuBr3}
\end{document}